\documentclass[12pt,english]{elsarticle}
\usepackage{mathrsfs}
\usepackage{amsmath}
\usepackage{amssymb}
\usepackage[T1]{fontenc}
\usepackage[latin9]{inputenc}
\usepackage{graphicx}
\graphicspath{ {Plots/} }
\usepackage{float}
\makeatletter

\usepackage[vlined,plain,noresetcount]{algorithm2e} 
\usepackage{colortbl}
\SetAlgoSkip{smallskip}
\usepackage{multicol}

\makeatother

\usepackage{babel}
\makeatletter

\RequirePackage{fix-cm}

\usepackage{todonotes}

\usepackage{transparent}

\newtheorem{theorem}{Theorem}
\newtheorem{definition}{Definition}

\newdefinition{remark}{Remark}
\newdefinition{corollary}{Corollary}
\newproof{proof}{Proof}
\newproof{thesis}{Thesis}

\usepackage{tikz}

\makeatother

\usepackage{babel}
\begin{document}

\title{Satisfaction of the Condition of Order Preservation: A Simulation Study}

\author{Jiri Mazurek}

\address{School of Business Administration in Karvina, Czech Republic}

\author{Konrad Ku\l akowski}

\address{AGH University of Science and Technology, Poland}

\begin{abstract}
We examine satisfaction of the condition of order preservation (COP) with respect to different levels of inconsistency for randomly generated multiplicative pairwise comparison matrices (MPCMs) of the order  $ n = \lbrace 3,4,...,9\rbrace $,  where a priority vector is derived both by the eigenvalue method (EV) and the geometric mean (GM) method. Our results suggest the GM method and the EV method preserve the COP condition almost identically, both for the less inconsistent matrices (with Saaty's consistency index $ CI <0.10 $), and the more inconsistent matrices (with $ CI \geq 0.10 $). Further, we find that the frequency of the COP violations grows  (almost linearly) with increasing  inconsistency of MPCMs measured by Koczkodaj's inconsistency index and Saaty's consistency index respectively, and we provide graphs to illustrate these relationships for MPCMs of the order $ n = \lbrace4, 7, 9\rbrace $. 
\end{abstract}
\begin{keyword}
pairwise comparisons\sep eigenvalue method\sep geometric mean method \sep condition of order preservation \sep numerical simulation.
\end{keyword}
\maketitle

\section{Introduction }

Pairwise comparisons belong among the oldest methods of decision making involving a selection of the best object (alternative, criterion, option, etc.) from a finite and non-empty set of objects. Since the number of pairwise comparisons grows as $ O(n^{2}) $, where $ n $ denotes the number of objects compared, real-world pairwise comparisons tasks are usually performed for relatively small number of objects. That is why in our study we focus on pairwise comparisons with $ n < 10 $.

The standard notion of consistency in (multiplicative) pairwise comparisons demands that if an object A is 4 times better than an object B, which in turn is 2 times better than an object C, then the object A must be exactly 8 times better than C. This can be considered a "numerical consistency". 

In 2008, the condition of order preservation (COP) in the context of pairwise comparisons was introduced in \citep{BanaCosta2008}. The COP states that after comparing a set of objects pairwise, the priority vector (weights associated with every compared object) should not contradict individual judgments. That is if an object A is directly preferred to an object B, then also the weight of A should be higher than the weight of B. Further, if A is compared to B, and C is compared to D, the difference between A and B being greater than the difference between C and D, then also the difference of weights associated with A and B should be greater than the difference between C and D.

This approach to consistency can be labelled as a "preferential consistency" (see section 3 for more precise definition).    
After its introduction, the condition of order preservation attracted attention of several authors, see e.g. \citep{Cavallo2020}, \citep{Kulakowski2015, Kulakowskietal2019}, \citep{MazurekRamik2019}. 

Studies \citep{Kulakowski2015, Kulakowskietal2019} provide a sufficient condition for the COP satisfaction with respect to inconsistency expressed by Koczkodaj's inconsistency index both for multiplicative case and more general case based on Alo-groups. The paper \citep{MazurekRamik2019} propose a new method for derivation of the priority vector based on the COP condition. 

However, little is known about how frequently is the COP condition satisfied for the eigenvalue (EV) method or the geometric mean (GM) method with respect to different inconsistency levels of pairwise comparisons, since up to date, there is no numerical study on the COP condition published in the literature. 

Therefore, the main objective of the paper is to investigate satisfaction/violation of the COP condition with respect to different levels of inconsistency for randomly generated multiplicative pairwise comparison matrices of the order  $ n = \lbrace 3,4,...,9\rbrace $ , where the priority vector is derived both by the EV and the GM method, in order to determine which method is better in the sense that it preserves the COP condition to a larger extent. 

Other objectives of the study include the examination of the relationship between the COP violation and the level of inconsistency in general, and the relationship between the COP satisfaction and the number of cases guaranteed to be satisfied by two theorems provided in section 3. 

Inconsistency of pairwise comparisons is measured via Koczkodaj's inconsistency index ($ KI $) and Saaty's consistency index ($ CI $) respectively.

The paper is organized as follows: section 2 provides preliminaries on pairwise comparisons, in section 3 the condition of order preservation (COP) is introduced, section 4 describes simulation procedure and it is followed by simulations results in section 5.

\section{Preliminaries }

The input data for the \emph{PC} method is a \emph{PC} matrix $C=[c_{ij}]$, where $c_{ij}\in\mathbb{R}_{+}$ and $i,j\in\{1,\ldots,n\}$. The values of $c_{ij}$ and $c_{ji}$ indicate the relative importance (or preference) of objects $a_{i}$ and $a_{j}$. 

\begin{definition}
\label{def:A-matrix-recip}A matrix $C$ is said to be (multiplicatively) reciprocal if:
\begin{equation}
 \forall i,j\in\{1,\ldots,n\}:c_{ij}=\frac{1}{c_{ji}} 
\end{equation}

and $C$ is said to be (multiplicatively) consistent if:
\begin{equation}
 \forall i,j,k\in\{1,\ldots,n\}:c_{ij}\cdot c_{jk}\cdot c_{ki}=1.
\end{equation}
\end{definition}

Since the \emph{PC} matrix contains subjective judgments of (human) epxerts, the condition (2) is usually not satisfied. 
This fact led to introduction of various inconsistency indices and studies of their properties, see e.g. \citep{Aguaron2003},\citep{Barzilai1998, Brunelli2013, BrunelliFedrizzi2015, Brunelli2017, Brunelli2018}, \citep{Golden1989, Koczkodaj1993}, or \citep{Mazurek2018}.

Perhaps the best known inconsistency indices are Saaty's consistency index $ CI $ and consistency ratio $ CR $ respectively, see \citep{Saaty1977, Saaty1980}, and Koczkodaj's inconsistency index $ KI $, see \citep{Koczkodaj1993}:

\begin{definition}
The eigenvalue based consistency index \emph{(Saaty's index)} of $n\times n$
reciprocal matrix $C$ is equal to: 
\begin{equation}
\textit{CI}=\frac{\lambda_{\textit{max}}-n}{n-1}\label{eq:Consistency_Index_AHP}
\end{equation}

where $\lambda_{\textit{max}}$ is the principal eigenvalue of $C$.
\end{definition}

The value $\lambda_{\textit{max}}\geq n$, and $\lambda_{\textit{max}}=n$
only if $C$ is consistent \cite{Saaty1980}. 

\begin{definition}
\label{def:Koczkodaj's-inconsistency-index}Koczkodaj's inconsistency
index $KI$ of $n\times n$ and ($n>2)$ reciprocal matrix $C=[c_{ij}]$
is equal to:
\begin{equation}
\textit{KI}=\underset{i,j,k\in\{1,\ldots,n\}}{\max}\left\{ \min\left\{ \left|1-\frac{c_{ij}}{c_{ik}c_{kj}}\right|,\left|1-\frac{c_{ik}c_{kj}}{c_{ij}}\right|\right\} \right\} \label{eq:1-koczkod_inc_idx-2}
\end{equation}
\end{definition}

\begin{remark}
From Definition 3 it follows that $ 0\leq KI < 1$. 
\end{remark}

The result of the pairwise comparisons method is a priority vector (vector of weights) $ w $.
According to the eigenvalue method proposed by Saaty \citep{Saaty1977}, vector $w$ is determined
as the rescaled principal eigenvector of $C$. Thus, assuming that
$Cw_{max}=\lambda_{\textit{max}}w_{max}$ the priority vector $w$
is 
\[
w=\gamma\left[w_{\textit{max}}(a_{1}),\ldots,w_{\textit{max}}(a_{n})\right]^{T}
\]

where $\gamma$ is a scaling factor. Usually it is assumed that $\gamma=\left(\sum_{i=1}^{n}w_{\textit{max}}(a_{i})\right)^{-1}$. 

According to the geometric mean method \citep{Crawford1987} 
the weight of i-th alternative is given by the geometric mean of the i-th row
of $C$. Thus, the priority vector is given as:
\begin{equation}
w=\gamma\left[\left(\prod_{r=1}^{n}c_{1r}\right)^{\frac{1}{n}},\ldots,\left(\prod_{r=1}^{n}c_{nr}\right)^{\frac{1}{n}}\right]^{T}\label{eq:GMM_def}
\end{equation}

where $\gamma$ is a scaling factor.

\section{Condition of Order Preservation (COP)}

The condition of order preservation (COP) was introduced in \citep{BanaCosta2008}:

\begin{definition}
\label{D4}
Let $\mathbf{C}=[c_{ij}]$ be a pairwise comparison matrix, and let $\mathbf{w} = (w_1,...,w_n)$ be a priority vector associated to $\mathbf{C}$. A PC matrix $\mathbf{C}$ is said to satisfy \emph{preservation of order preference condition (POP condition)}  with respect to priority vector $\mathbf{w}$ if
\begin{equation}
c_{ij}>1 \Rightarrow w_i > w_j. \label{8}
\end{equation}
\end{definition}

\begin{definition}
\label{D5}
Let $\mathbf{C}=[c_{ij}]$ be a pairwise comparison matrix, and let $\mathbf{w} = (w_1,...,w_n)$ be a priority vector associated to $\mathbf{C}$. A PC matrix  $\mathbf{C}$ is said to satisfy \emph{ preservation of order of intensity of preference (POIP condition)}  with respect to vector $\mathbf{w}$ if 
\begin{equation}
c_{ij}>1, c_{kl}>1,\text{ and } c_{ij}>c_{kl} \Rightarrow \frac{w_i}{w_j} > \frac{w_k}{w_l} . \label{9}
\end{equation}
\end{definition}

\begin{remark}
In previous definitions it is required that relations (6) and (7) are satisfied for all pairs of indices $ (i,j) $ and all quadruples of indices $ (i,j,k,l) $ respectively. In general, there are $ (n^{2}-n) $ individual POP conditions and $ (n^{2}-n)(n^{2}-n-2) $ individual POIP conditions.
\end{remark}

The following two theorems, see \citep{Kulakowski2015}, \citep{Kulakowskietal2019} provide sufficient conditions to satisfaction of the POP and POIP conditions respectively. 

\begin{theorem}
For the PC matrix $ \mathbf{C}(c_{ij}) $ with Koczkodaj's inconsistency index KI and the ranking vector $ w $ obtained by the EV or GM method holds that:
\begin{equation}
c_{ij} > \frac{1}{1-KI} \Rightarrow w(a_{i}) > w(a_{j})
\end{equation}
\end{theorem}

\begin{theorem}
For the PC matrix $ \mathbf{C}(c_{ij}) $ with Koczkodaj's inconsistency index KI and the ranking vector $ w $ obtained by the EV or GM method holds that:
\begin{equation}
\frac{c_{ij}}{c_{kl}} > (\frac{1}{1-KI})^{2} \Rightarrow \dfrac{w(a_{i})}{w(a_{j})} > \dfrac{w(a_{k})}{w(a_{l})}
\end{equation}
\end{theorem}

\section{Numerical Simulations }

Numerical simulations were performed to examine two main problems:
\begin{itemize}
\item How frequently are individual POP and POIP conditions met for different levels of inconsistency with respect to the EV and the GM method. 
\item How often are individual POP and POIP conditions met for different levels of inconsistency with respect to Theorems 1 and 2. 
\end{itemize}
Simulations were performed for pairwise comparisons matrices \textit{C} of the order $ n = \lbrace3,4,5,6,7, 8, 9 \rbrace $.

In the beginning, a random pairwise comparisons matrix $n\times n$ was created as follows:
\begin{enumerate}
\item In the first step the vector $w=[w(a_{1}),\ldots w(a_{n})]^{T}$ was randomly drawn, where every $w(a_{i})\in[1,9]$,
\item Then, the fully consistent matrix $C=\left[\frac{w(a_{i})}{w(a_{j})}\right]$ was created.
\item Next, for the given disturbance level $1<\gamma<4$ for every entry of $C$ the actual disturbance coefficient $\delta$ was drawn such that $\delta\in[1/\gamma,\gamma]$ .
\item In the next step the matrix $C$ was altered to $\widetilde{C}=\left[\frac{w(a_{i})}{w(a_{j})}\cdot\delta_{ij}\right]$
where $\delta_{ij}\in[1/\gamma,\gamma]$, and for each pair $\{i,j\}$
the value $\delta_{ij}$ was chosen separately. 
\item The matrix $\widetilde{C}$ was evaluated for the POP and POIP conditions and inconsistency (KI and CI indices) by the EVM or GMM.
\end{enumerate}

For every matrix size $3\times3$ to $9\times9$ and every disturbance level, 100 or 500 random matrices were generated (210,000 or 1,050,000 matrices respectively in total). 

The results have a form of an Excel file and can be freely downloaded from Mendeley data storage \citep{data}.  


\section{Results}

Table 1 provides average percentage of satisfied individual POP and POIP conditions for multiplicative pairwise comparison matrices of the order $ 3 \leq n \leq 9 $, when the priority vector (vector of weights) was elicited via EV method and GM method, whilst $ CI < 0.10 $, that is this case relates to matrices sufficiently consistent according to Saaty \citep{Saaty1980}. It can be seen that individual POP conditions were satisfied in approximately 90\% of all cases, and individual POIP conditions in approximately 96\% of all cases. 

Table 2 provides analogous data for matrices with $ CI \geq 0.10 $, that is matrices considered by Saaty \citep{Saaty1980} to be too inconsistent. It's clear that these less consistent matrices are worse in satisfaction of the POP and POIP conditions by several percent. 

In addition, Theorems 1 and 2 are more efficient (they capture more cases) for matrices with $ CI < 0.10 $ and $ n = \lbrace 3,4\rbrace $, see the last two columns in Tables 1 and 2.  

As for the comparison between EV method and GM method, it is evident that both methods yield almost identical results. 

The relationship between average number of violated POP and POIP individual conditions with respect to Koczkodaj's index $ KI $ and  
$ n = \lbrace4,7,9\rbrace $ and EV method is shown in Figures 1-6.
It can be seen that the number of individual violations of the POP and POIP conditions grows with increasing KI roughly linearly.

Figures 7-10 show average numbers of matrix entries satisfying POP and POIP individual conditions via Theorems 1 and 2 with respect to KI and EV method for $ n = \lbrace4,7\rbrace $. Efficiency of Theorems 1 and 2 decreases approximately linearly (for POP) and hyperbolically (for POIP) with the growing $ KI $.

\begin{center}
\begin{table}

\caption{Table 1. Satisfaction of POP and POIP conditions (in \%), $ CI < 0.10 $. EV method was used for the last two columns}.
\\
\begin{tabular}{ |c|c|c|c|c|c|c| } 
\hline
n & POP (EV) & POP (GM) & POIP (EV) & POIP (GM) & POP Th1 & POIP Th2 \\
\hline 
3 & 91.02 & 91.02 & 96.60 & 96.60 & 58.25 & 7.60\\ 
4 & 90.44 & 90.59 & 95.79 & 95.79 & 40.39 & 3.50\\ 
5 & 89.70 & 89.88 & 95.80 & 95.82 & 32.90 & 2.53\\ 
6 & 89.91 & 90.01 & 95.80 & 95.82 & 29.41 & 2.25\\ 
7 & 89.66 & 89.74 & 95.83 & 95.85 & 27.10 & 2.12\\ 
8 & 89.57 & 89.64 & 95.98 & 96.00 & 26.45 & NA\\
9 & 89.62 & 89.70 & 96.03 & 96.05 & 25.28 & NA\\
\hline
\end{tabular}
\label{table:1}
\end{table}
\end{center}

\begin{center}
\begin{table}
\caption{Table 2. Satisfaction of POP and POIP conditions (in \%),  $ CI \geq 0.10 $. EV method was used for the last two columns}.
\\
\begin{tabular}{ |c|c|c|c|c|c|c| } 
\hline
n & POP (EV) & POP (GM) & POIP (EV) & POIP (GM) & POP Th1 & POIP Th2 \\
\hline 
3 & 87.33 & 87.33 & 96.00 & 96.00 & 45.83 & 5.29\\ 
4 & 85.88 & 86.29 & 94.01 & 94.03 & 21.30 & 0.53\\ 
5 & 83.64 & 84.21 & 93.63 & 93.69 & 9.72 & 0.05\\ 
6 & 82.70 & 83.06 & 93.56 & 93.64 & 5.45 & 0.01\\ 
7 & 82.01 & 82.55 & 93.43 & 93.50 & 3.27 & 0.00\\ 
8 & 81.71 & 82.18 & 93.48 & 93.55 & 2.22 & NA\\
9 & 81.37 & 81.83 & 93.40 & 93.46 & 1.52 & NA\\
\hline
\end{tabular}

\label{table:2}
\end{table}
\end{center}

\begin{figure}[H]
\centering
\includegraphics[scale=0.49]{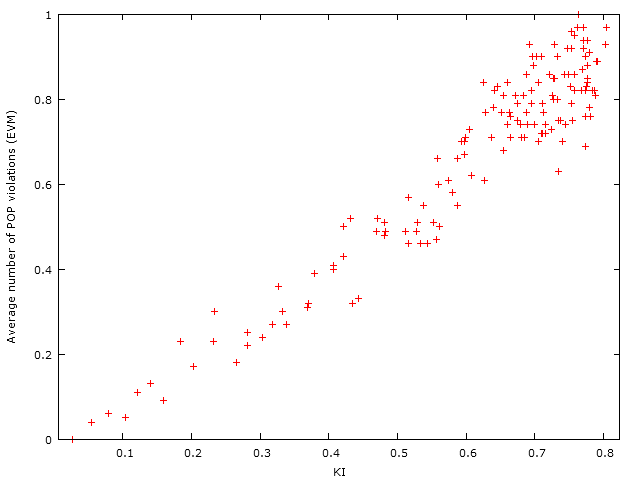}
\caption{Average number of individual POP violations (EVM), n = 4}
\end{figure}

\begin{figure}[H]
\centering
\includegraphics[scale=0.49]{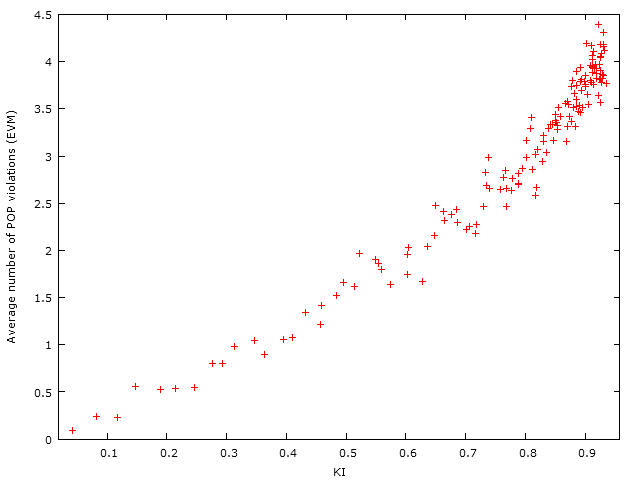}
\caption{Average number of individual POP violations (EVM), n = 7}
\end{figure}

\begin{figure}[H]
\centering
\includegraphics[scale=0.49]{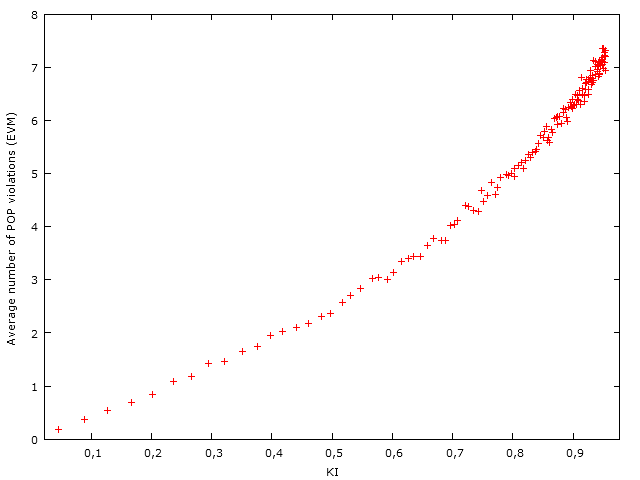}
\caption{Average number of individual POP violations (EVM), n = 9}
\end{figure}

\begin{figure}[H]
\centering
\includegraphics[scale=0.49]{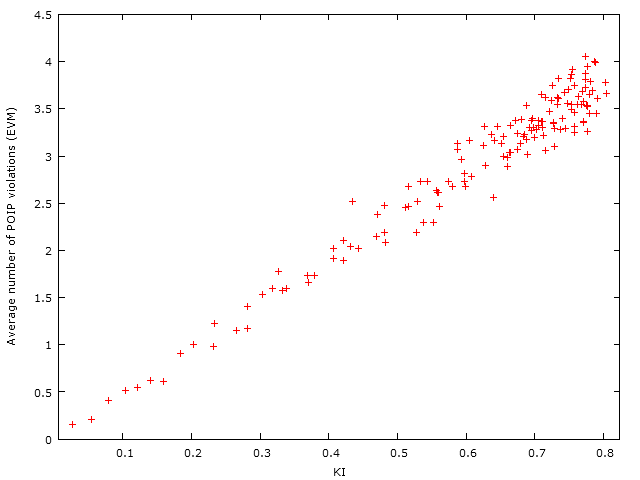}
\caption{Average number of individual POIP violations (EVM), n = 4}
\end{figure}

\begin{figure}[H]
\centering
\includegraphics[scale=0.49]{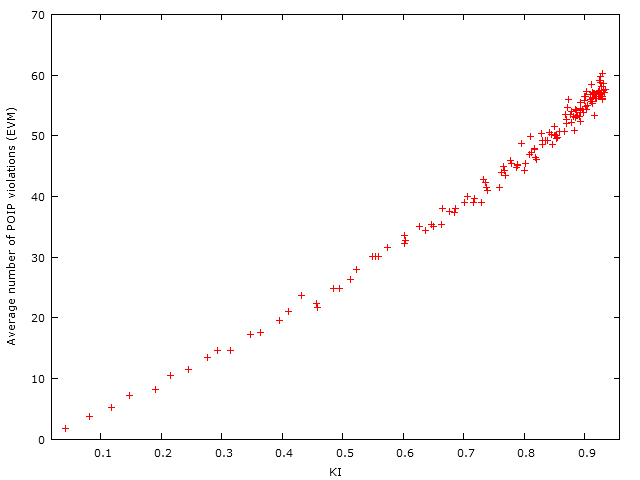}
\caption{Average number of individual POIP violations (EVM), n = 7}
\end{figure}

\begin{figure}[H]
\centering
\includegraphics[scale=0.49]{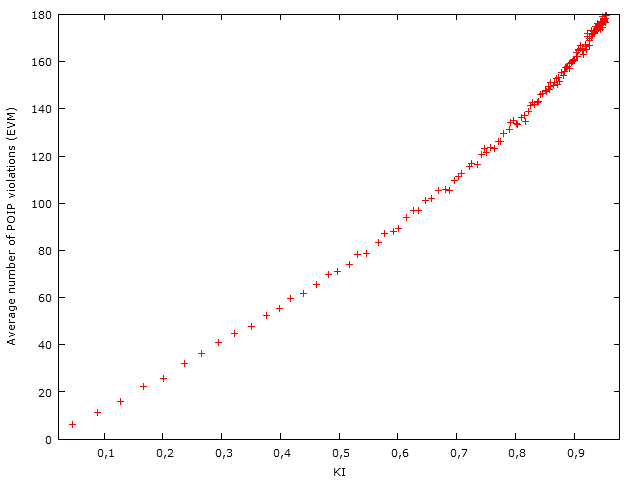}
\caption{Average number of individual POIP violations (EVM), n = 9}
\end{figure}

\begin{figure}[H]
\centering
\includegraphics[scale=0.49]{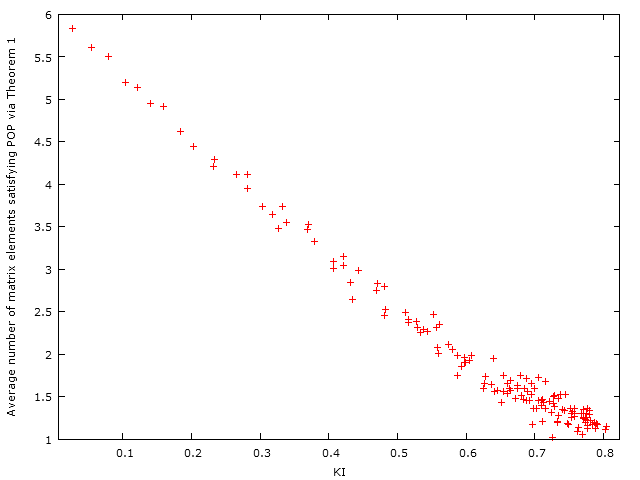}
\caption{Average number of inidvidual POP satisfied via Theorem 1 (EVM), n = 4}
\end{figure}

\begin{figure}[H]
\centering
\includegraphics[scale=0.49]{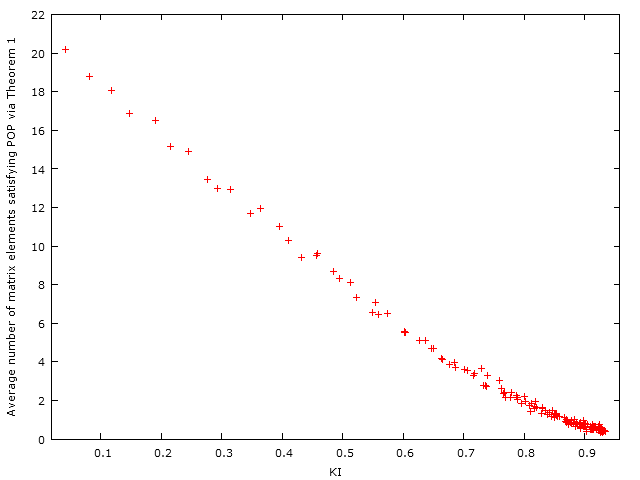}
\caption{Average number of individual POP satisfied via Theorem 1 (EVM), n = 7}
\end{figure}

\begin{figure}[H]
\centering
\includegraphics[scale=0.49]{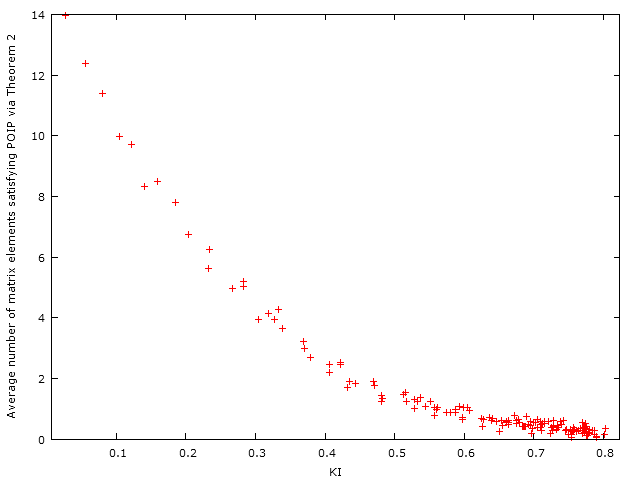}
\caption{Average number of individual POIP satisfied via Theorem 2 (EVM), n = 4}
\end{figure}

\begin{figure}[H]
\centering
\includegraphics[scale=0.49]{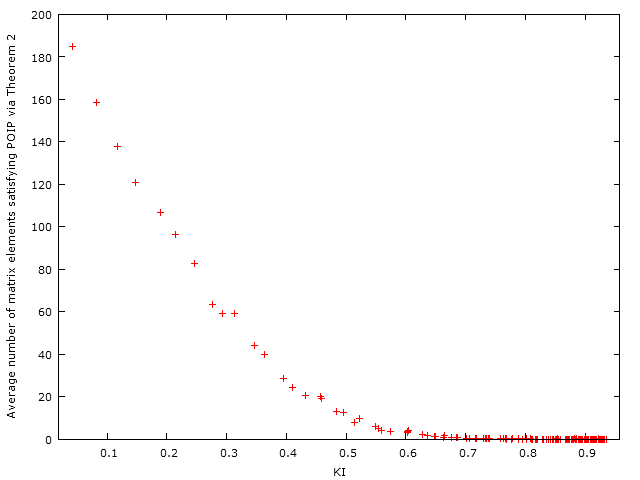}
\caption{Average number of individual POIP satisfied via Theorem 2 (EVM), n = 7}
\end{figure}


\section{Conclusions}
The aim of the paper was to investigate how frequently is condition of order preservation (COP) satisfied or violated for randomly generated multiplicative pairwise comparison matrices up to the order of $ n = 9 $ with respect to their inconsistency expressed by Koczkodaj's inconsistency index and Saaty's consistency index for both EV and GM method.

Numerical results suggest that the number of violations of individual POP and POIP conditions grows with the growing inconsistency, and the geometric mean method and the eigenvalue method yield almost identical results, therefore, for practical purposes, the priorization method in the COP framework does not matter. 
 
Also, it was found that the average number of individual POP conditions guaranteed by Theorem 1 decreases roughly linearly, and the average number of individual POIP conditions guaranteed by Theorem 2 decreases roughly hyperbolically. Both theorems were significantly more efficient (captured significantly more cases) for smaller matrices (of the order $ n = \lbrace3,4\rbrace $ and less inconsistent matrices (with $ CI < 0.10 $), than for larger ($ n \geq 5) $ and more inconsistent matrices (with $ CI \geq 0.10) $). 

Further research may examine other aspects for the COP condition, such as the maximum of possible individual violations, or the relationship between "numerical consistency" and "preferential consistency" in general. 


\section*{Acknowledgment}
The research was supported by the project GACR Nr.~18-01246S.

\end{document}